\documentclass[aps,prl,twocolumn,showpacs,superscriptaddress,groupedaddress]{revtex4}  
\usepackage{graphicx}  
\usepackage{dcolumn}   
\usepackage{bm}        
\usepackage{amssymb}   
\usepackage{amsmath}
\usepackage{color}

\def\be{\begin{equation}}
\def\ee{\end{equation}}
\def\bea{\begin{eqnarray}}
\def\eea{\end{eqnarray}}

\begin{document}

\widetext


\title{Tetraquarks in the $1/N$ Expansion: a New Appraisal}
\author{Luciano Maiani}
\email{Luciano.Maiani@cern.ch}
\author{Antonio D. Polosa}
\email{antonio.polosa@roma1.infn.it}
\author{Veronica Riquer}
\email{veronica.riquer@cern.ch}
\affiliation{Dipartimento di Fisica and INFN,  Sapienza  Universit\`a di Roma, Piazzale Aldo Moro 2, I-00185 Roma, Italy.}
\date{\today}

\begin{abstract}
We discuss the necessary, albeit not sufficient, conditions for tetraquark poles to occur in the $1/N$ expansion of QCD and find the minimum order  at which such poles may appear. 
Assuming tetraquark poles, we find a new non-planar solution with the minimal number of topologies and tetraquark species. The solution implies narrow states. Mixing with quarkonium states is allowed so that $P$-wave tetraquarks with $J^{PC}=1^{--}$ would couple to $e^+e^-$. 
\end{abstract}

\maketitle
 
 \section{Introduction and Presentation of the Results}\label{intro}
 QCD attraction between two quarks in color antisymmetric configuration has led to speculations about a tetraquark (diquark-antidiquark) interpretation of light scalar mesons~\cite{Jaffe:2003sg}, of hidden charm $X,~Y,~Z$ ~\cite{Maiani:2004vq,Maiani:2014aja} and hidden beauty $Z_b$~\cite{Ali:2011ug} hadrons. 
 Double charm and beauty tetraquarks have been considered in~\cite{Esposito:2013fma,Guerrieri:2014nxa} and, more recently, in~\cite{Karliner:2017qjm} and~\cite{Eichten:2017ffp,Eichten:2017ual}. In the latter papers, in particular, it is pointed out that  the attraction in the $bb$ diquark may be so strong as to make the double $b$ tetraquark to be stable under strong interaction decays. The presence of tetraquarks in the double charm or beauty channel is also indicated in a number of non-perturbative approaches such as the Heavy Quark-Diquark Symmetry~\cite{Manohar:2000dt,Savage:1990di,Brambilla:2005yk,Mehen:2017nrh},  lattice QCD, see e.g.~\cite{Bicudo:2016ooe,Francis:2016hui}
and references therein, and the Born-Oppenheimer approximation, see~\cite{Bicudo:2017szl} and references therein.
 
 In the present, paper we study tetraquarks in the $1/N$ expansion of QCD~\cite{ninfty,Witten:1979kh} following investigations in~ \cite{Coleman:1985},~\cite{Weinberg:2013cfa}, ~\cite{knecht,Lebed:2013aka,Esposito},~\cite{Cohen:2014tga},~\cite{Maiani:2016hxw} and~\cite{Lucha:2017mof}. Our  aim is to identify the lowest order planar diagrams in the $1/N$ expansion fulfilling the {\it necessary conditions} for tetraquark poles to appear.  We stress that, as was the case of previous analyses~\cite{Weinberg:2013cfa,knecht,Lebed:2013aka,Esposito,Cohen:2014tga,Maiani:2016hxw,Lucha:2017mof}, these conditions are not sufficient to guarantee the existence of the pole. Indeed a positive proof of the existence of tetraquark poles in meson-meson correlation functions is not available at present. 
 
 The existence of tetraquarks is being tackled with different non perturbative methods, such as lattice gauge theory. In the case of $b$ quarks, one may wonder if the large quark mass and large $N$ limits are interchangeable. Present lattice calculations are based either on dynamical heavy quarks with fixed mass, in the case of charm, or  Non-Relativistic QCD. Both approaches can be included in large $N$ QCD (see e.g.~\cite{Witten:1979kh} for an explict non-relativistic treatment of heavy quarks in large $N$ QCD) and we feel that a comparison of ours with lattice results may be useful. 
  
In~\cite{Weinberg:2013cfa,knecht,Lebed:2013aka,Esposito} it was assumed that a tetraquark pole could appear in meson-meson scattering to the first non leading order in $1/N$, obtaining~\cite{Weinberg:2013cfa,knecht,Lebed:2013aka,Esposito} 
\be
 g_{T M_1 M_2}\propto \frac{1}{\sqrt{N}}\label{decamp0},~ {\rm (first~non~leading~order)}
\ee 
for the tetraquark-meson-meson coupling. 

The assumption of tetraquarks in diagrams of leading order was criticised in~\cite{Cohen:2014tga,Maiani:2016hxw} on the basis that the four quark cuts in the diagrams considered always corresponded to two non interacting mesons, and it was not reasonable to assume these configurations to form a tetraquark, even after including the full non-perturbative corrections permitted by the given order in $1/N$. Both papers suggested that higher order diagrams in the $1/N$ expansion had to be considered.  

\begin{figure}[htb!]
\begin{center}
\includegraphics[width=7truecm]{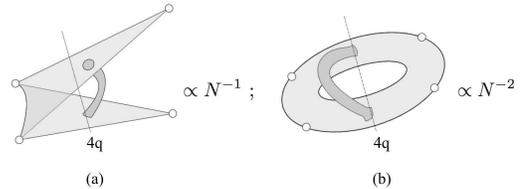}
 \end{center} 
 \caption{\footnotesize Meson-meson scattering amplitudes with four-quarks intermediate states in~\cite{Maiani:2016hxw}. The handle is a necessary condition for the existence of genuine tetraquark cuts. Gray shaded surfaces indicate the full set of planar gluon interactions}
\label{handle1}
\end{figure}

It was noted in Ref.~\cite {Maiani:2016hxw} that for diagrams with a four quarks cut the additional interaction between the $q\bar q$ pairs to form a tetraquark requires at least one  handle. The analysis was based on the two diagrams of Fig.~\ref{handle1} that, saturating with one tetraquark pole, give:    
 \bea
 && g_{T M_1 M_2}\propto \frac{1}{N\sqrt{N}},~{\rm Fig.~\ref{handle1}(a)}\label{decamp1} \\
 && f(T\leftrightarrow c\bar c)\propto \frac{1}{N\sqrt{N}},~{\rm Fig.~\ref{handle1}(b)}\label{mix0} 
\eea
where (\ref{mix0}) gives the tetraquark-charmonium mixing amplitude.

\begin{figure}[htb!]
 \begin{center}
\includegraphics[width=7truecm]{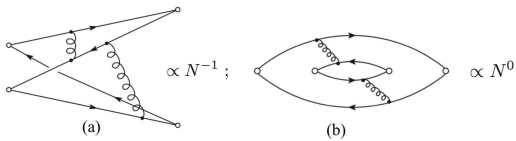}
\end{center} \caption{\footnotesize  Meson-meson scattering with four-quarks intermediate states in~\cite{Lucha:2017mof}. $(a)$  is topologically equivalent to Fig.~\ref{handle1}(a) but $(b)$ has no handles. Quark bilinear sources in $(b)$ are inserted on both external and internal quark loops.} 
 \label{lucha1}
\end{figure}

Futher progress was made in Ref.~\cite{Lucha:2017mof} where the existence of non trivial Landau singularities in the $s$- channel is adopted as a criterion to identify candidate diagrams for a tetraquark pole.  This criterion leads the authors to the diagrams in Fig.~\ref{lucha1}. 

Fig.~\ref{lucha1}(a)  is a perturbative representation of  Fig.~\ref{handle1}(a) (thus equivalent to it) but the diagram in~\ref{lucha1}(b) introduces a novel feature, with respect to Fig.~\ref{handle1}(b), in that the insertions in the internal loop may correspond to bilinears with unequal quark flavours, like in Figs.~\ref{handle1}(a)  and \ref{lucha1}(a). Thus, the same tetraquark(s) may appear as intermediate states in both diagrams, unlike what happens in Fig.~\ref{handle1}. This result conflicts with the different order in the $1/N$ expansion of the two diagrams. The authors propose to overcome the problem by assuming two distinct tetraquarks with the same flavour content~\cite{Lucha:2017mof}.   

In the present paper, we argue that the presence of Landau singularities is not enough. As we shall see in the next Section, the intermediate state in the diagram of Fig.~\ref{lucha1}(b), even after non-perturbative addition of all planar gluons between each $q\bar q$ pair, can still be made by two non-interacting pairs and not be  able to develop a tetraquark pole. 

\begin{figure}[htb!]
 \begin{center}
   \includegraphics[width=7truecm]{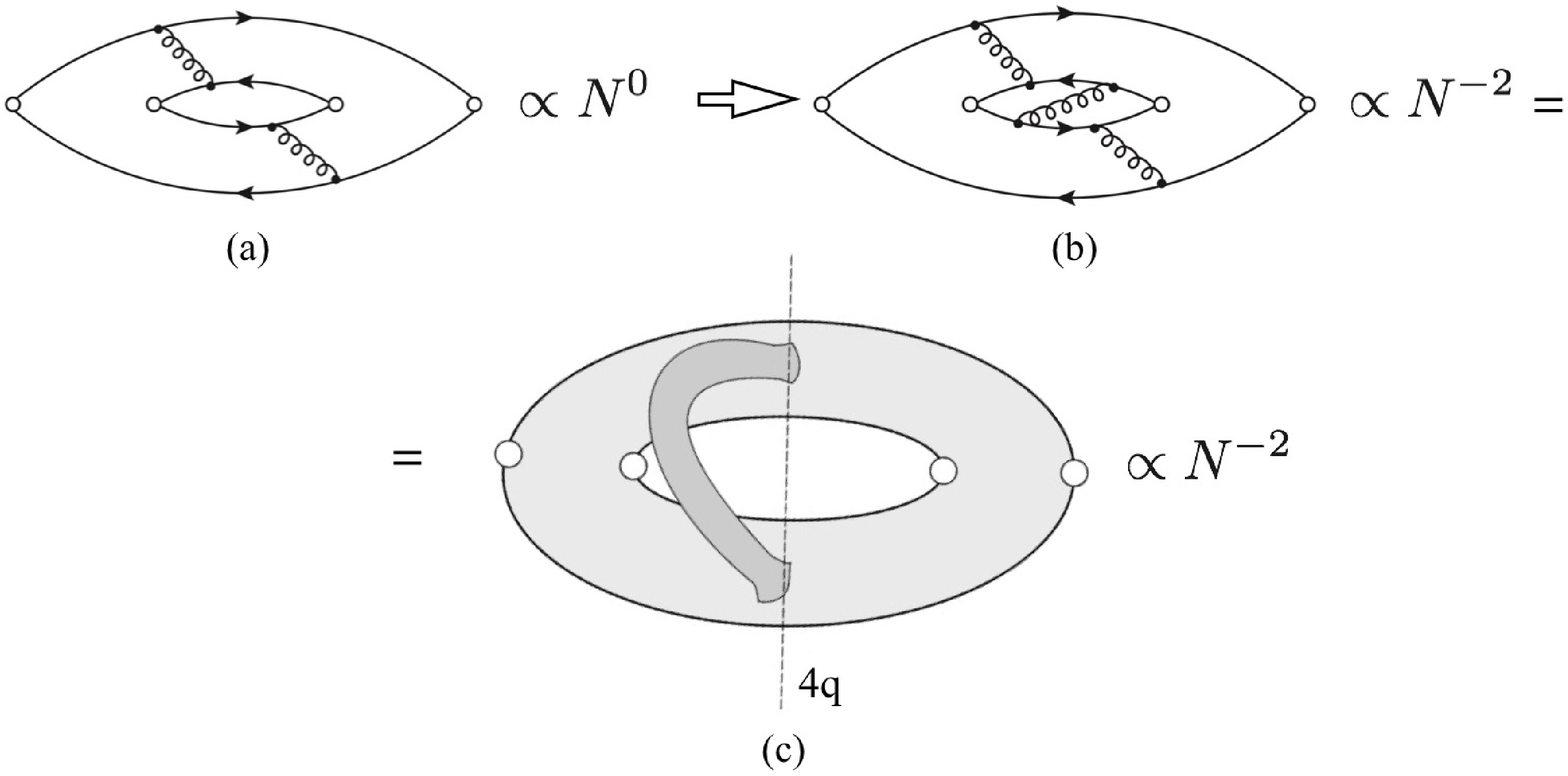}
 \end{center} \caption{\footnotesize (c) An handle is added to the diagram in Fig.~2(b), changing the order in the $1/N$ expansion. In this way, we recover the diagram in Fig.~1(b), with a different insertions of the quark bilinear sources.}
 \label{luchamod}
\end{figure}

All the planar gluons are effectively correlating the quark pairs in any cut but an interaction between the two pairs in the cut, is not guaranteed. We seek a solution which enforces unambiguously this interaction, as it occurs  by introducing all shortcut non-planar gluons summarized by the 
non-perturbative diagram with one handle of Fig.~\ref{luchamod}(c). 
 
This reduces the order from $N^0$ to $N^{-2}$, see Fig.~\ref{luchamod}, due to the addition of one handle to the otherwise planar diagram of Fig.~\ref{lucha1}(b).

Saturation with unequal flavour tetraquarks now must be considered in both diagrams of Figs.~\ref{handle1}(a) and~\ref{luchamod}(c). The two diagrams differ again in the order of the $1/N$ expansion. However the order is now inverted and the solution of having two different tetraquarks as in  ~\cite{Lucha:2017mof} is not available.

In fact, denoting by $\alpha$ the order~ in ~$N$ of a given diagram~proportional to $N^\alpha$,  we shall see that saturation of the two correlation functions with any number of tetraquarks implies that:
\be
\alpha \left[{\rm Fig.}~\ref{luchamod}(c)\right] \geq \alpha \left[{\rm Fig.}~\ref{handle1}(a)\right]\label{ineqorder}
\ee
while the two orders are $-2$ and $-1$, respectively~\footnote{The correlation functions in Fig.~\ref{lucha1} do satisfy the inequality~(\ref{ineqorder}), which is why two tetraquarks work in Ref.~\cite{Lucha:2017mof}. However, as just indicated, the correlator of Fig.~\ref{lucha1}(b) to order $N^0$ does not lead to any tetraquark pole. With the tetraquark pole in Fig.~\ref{lucha1}(a), one falls in the Option 2.}.

Consistency forces us to conclude that only one class of diagrams can develop the tetraquark pole, either  Figs.~\ref{handle1}(b) and \ref{luchamod}(c), or Figs.~\ref{handle1}(a) and ~\ref{lucha1}(a).  Thus we are led to two possible consistent solutions for the tetraquark couplings. Both solutions require one tetraquark only for given flavour configuration, consistent with what expected in QCD, where only the color antisymmetric quark-quark channel is attractive. 

\paragraph{{\bf Option 1.}} Tetraquark pole in  Figs.~\ref{handle1}(b) and \ref{luchamod}(c) leads to a novel solution with the following features. 
  \begin{itemize} 
 \item For given flavours, one tetraquark suffices;
\item tetraquark-meson-meson coupling:
 \begin{eqnarray}
&& g_{T M_1 M_2}\propto \frac{1}{N^2}\label{decampfin}
\end{eqnarray}
 \item the transitions: $Y\to Z+\pi$ and $Y\to X+\gamma$ are allowed:
 \begin{eqnarray}
 && A(T\to T^\prime +M)\propto \frac{1}{\sqrt{N}}\label{high}\\
&& A(T\to T^\prime +\gamma)\propto e N^0 \label{em}
\end{eqnarray}
\item mixing of neutral, hidden-charm tetraquarks with charmonia:
\be 
f(T\leftrightarrow c\bar c)\propto \frac{1}{N\sqrt{N}} 
\ee
\item $Y$ states may be produced by or annihilate into $e^+ e^-$ via mixing. 
 \end{itemize}
The result in Eq.~(\ref{high}) is the same found in Eq.(\ref{decamp0}), see~\cite{Weinberg:2013cfa,knecht}, and the conclusion that the width of tetraquarks vanishes for $N \to \infty$ still applies. 

Interestingly the solution allows for tetraquark-charmonium mixing, suggested by the coupling of putative tetraquarks, the $Y$-states with $J^{PC}=1^{--}$, to the $e^+e^-$ channel.

 \paragraph{\bf Option 2.}Tetraquark pole in Figs.~\ref{handle1}(a) and ~\ref{lucha1}(a) gives back the solution found in Ref.~\cite{Maiani:2016hxw} for tetraquark-meson-meson coupling, Eq.~(\ref{decamp1}), with the additional condition that the mixing parameter, previously given by (\ref{mix0}), has to vanish.
 \begin{figure}
\begin{center}
\includegraphics[width=0.80\columnwidth]{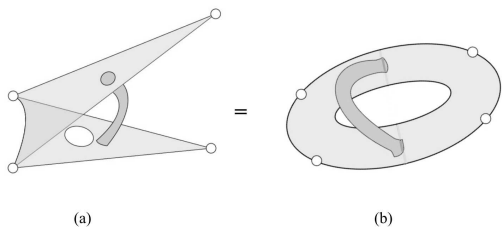}
 \end{center}
\caption{\footnotesize In (a) and (b) two equivalent drawings of diagram in Fig.~\ref{handle1}(a) with the addition of one fermion loop. }  
\label{rombo3}
\end{figure}

The presence of a mixing between tetraquarks and charmonia makes Option 1 phenomenologically more attractive.  

We cannot decide mathematically if  the diagram of Fig~\ref{handle1}(a), has or has not a tetraquark pole. 
We note, however, that the $s$-channel cut of the diagram in Fig.~\ref{rombo3}(a) corresponds to the $u$-channel cut of the diagram ~\ref{rombo3}(b) that is supposed, in Option 1, to have the tetraquark in its $s$ (and $t$) channel. No tetraquark in Fig.~\ref{rombo3}(a) would reproduce a situation similar to the one encountered in meson-meson scattering in the leading order amplitude, which has meson poles in the $s$ and $u$ channels, but not in the $u$ channel. Thus Option 1 is not so unreasonable.

\section{Tetraquarks in large $N$ QCD}
\label{largeN:sec}
 
We discuss here in detail why we think that the diagram in Fig.~\ref{lucha1}(b) cannot produce a tetraquark pole, even after the addition of all higher orders in the QCD coupling $g$ compatible with being of order $N^0$ in the $1/N$ expansion.

\begin{figure}
\begin{center}
\includegraphics[width=1.0\columnwidth]{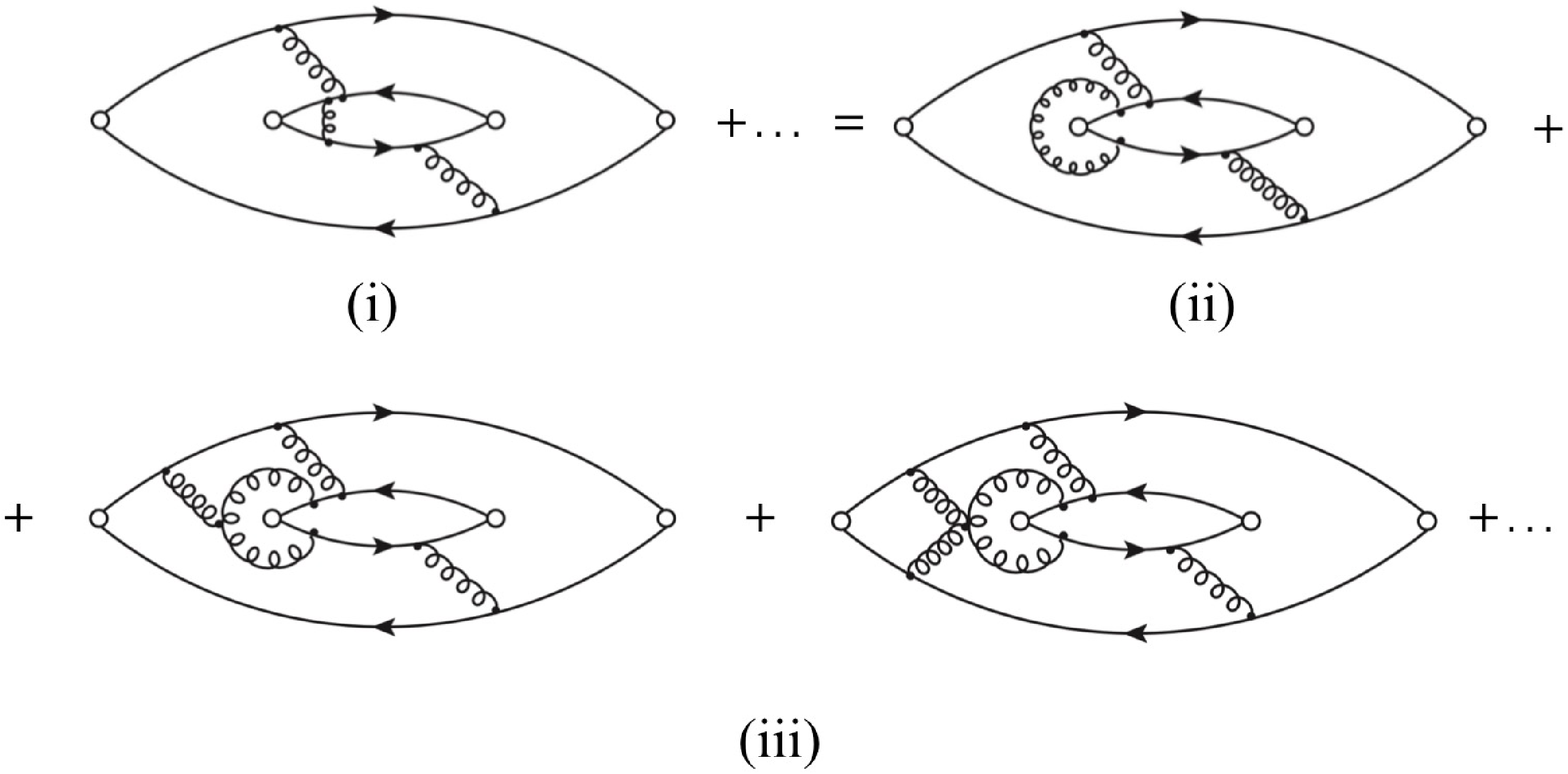}
 \end{center}
\caption{\footnotesize 
}  
\label{vertextot}
\end{figure}

In principle, one could think that by adding one gluon next to the bilinear insertion in the internal loop, as in Fig.~\ref{vertextot}(i), corresponds to a genuine interaction that leads, after iteration of similar corrections, to the binding of the $q\bar q$ pairs into a tetraquark.

For the diagram to be of order $N^0$, one must be able to redraw it as a planar diagram with gluons between the the two quark loops, which is done in Fig.~\ref{vertextot}(ii). However \ref{vertextot}(ii) does  not correspond to an interaction between the two $q\bar q$ pairs along any vertical cut one may draw. This conclusion is reinforced by  Fig~\ref{vertextot}(iii), which represents the first corrections of higher order in $g$ but still of order $N^0$. Diagrams (ii), (iii) and other higher order planar corrections simply determine the momentum and color flow from the external bilinear insertions into each $q\bar q$ pair, but produce no interaction among them. The four-fermion cut of diagrams (ii) and (iii) produces two $q\bar q$ pairs that are correlated in color, being both either in color singlet or in color octet. The correletion is a consequence of color conservation entirely analogous to the spin correlation in a pair of photons emitted in $J=0$. Nobody would attribute the correlation to a real interaction between the the two photons.

On the contrary, the intermediate gluon inserted in the internal loop of Fig.~\ref{luchamod}(b) cannot be fitted outside the loop in a planar way.

Such a non-planar gluon corresponds to a non-eliminable interaction between the $q\bar q$ pairs in the upper and lower decks of the diagram and it can be iterated at the same order in $N$. The non-planar gluon is the perturbative representation of a ``handle" set on an otherwise planar diagram, as in Fig.~\ref{luchamod}(c). The resulting amplitude is of order $N^{-2}$, as can be checked by explicit calculation.

In conclusion, we see no interaction until order $N^{-2}$.

\section{Couplings and mixing parameters} 

The order $N^\alpha$ of each diagram is computed from the 't Hooft rule~\cite{ninfty} 
 \be
 \alpha=2-L-2H
 \label{counting} 
 \ee
 where $L$ is the number of fermion loops and $H$ the number of handles. 
 
 For Fig.~\ref{handle1}(a): $L=1,~H=1$ and $\alpha=-1$;  for Figs.~\ref{handle1}(b),~\ref{luchamod}: $L=2,~H=1$ and $\alpha=-2$. 
 
 We consider meson-meson scattering  in the diagrams of Fig.~\ref{handle1}(a) and Fig.~\ref{luchamod}, starting with the case of all unequal flavors.

\paragraph{{\bf All unequal flavors}.} 
We have four quark-antiquark bilinears, {\it e.g.}: $M_1=M(b\bar u)$,
 $M_2=M(d\bar c)$, $M_3=M(b\bar c)$, $M_4=M(d\bar u)$, and two possible initial or final states:
\bea
&& A: M_1+M_2\nonumber \\
&& B: M_3+M_4
\eea
The meson-meson amplitude is a two-by-two matrix, the diagram in Fig.~\ref{handle1} (a) corresponds to non-diagonal transitions $A\to B$ and $B\to A$ and Fig.~\ref{luchamod} to diagonal ones: $A\to A$, $B\to B$. We define the couplings
\be
g_{_{A,B}}=A(T\to A,B)
\label{defsab}
\ee

If we saturate both diagrams with a single tetraquark $T(bd\bar u \bar c$) we obtain (with $\sqrt{N}$ the wave function normalization of each meson)
\bea
&&\left(\sqrt {N}\right)^4 g_{_A} g_{_B} \propto \frac{1}{N} \Rightarrow g_{_A} g_{_B} \propto N^{-3}\label{triang1}\\
&& \left(\sqrt {N}\right)^4 g_{_{A,B}}^2 \propto \frac{1}{N^2}\Rightarrow g_{_{A,B}}^2 \propto N^{-4}\label{triang2}
\eea
For large $N$ the  latter two relations do not satisfy the triangle inequality and are obviously incompatible with each other, no matter how many tetraquarks with the same flavor composition one assumes. 

We consider explicitly the case where  {\it the tetraquark pole appears only in Figs.~\ref{luchamod}(c) and~\ref{handle1}(b)} and (\ref{triang1}) does not apply, Option 1 of the Introduction.

By inserting a tetraquark in the $s$-channel of ~\ref{luchamod}(c), we obtain
\be
g_{_A}=g_{_B}=g \propto \frac{1}{N^2}.
\label{couplingfin}
\ee

\begin{figure}
 \begin{center}
   \includegraphics[width=0.90\columnwidth]{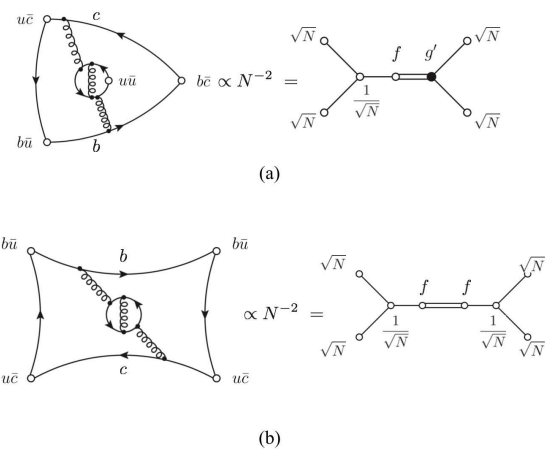}
 \end{center}
\caption{\footnotesize  With quark bilinears all inserted on the external fermion loop, diagram (a) represents another contribution to meson-meson scattering, in addition to diagram in Fig~\ref{luchamod}. The $s$-channel cuts correspond to the figure on the right and determine the mixing $q\bar q$-tetraquark. Diagram (b): the $s$-channel cuts, as indicated by the figure on the right, determine the product $fg$, in agreement with Eqs.~(\ref{couplingfin}) and~(\ref{mixing}).} 
\label{manypoles}
\end{figure}

Next, one may then consider  Fig.~\ref{handle1}(b) where meson insertions are all on the external quark loop. This introduces the mixing of a tetraquark with two equal flavours with $q\bar q$ mesons, Fig.~\ref{manypoles}(a), and one obtains the mixing parameter $f$:
\be
 f \propto \frac{1}{N\sqrt{N}}.
\label{mixing}   
\ee 

One may ask if a tetraquark pole can appear in the higher orders of the diagram \ref{handle1}(a). The next order in $1/N$ is obtained by adding one fermion loop and leads to the diagram of Fig.~\ref{rombo3}(a). 
Strictly speaking, this diagram has a six-quarks cut, not indicative of a tetraquark pole. 
As Fig.~\ref{rombo3} shows, however, the diagram \ref{rombo3}(a) is topologically equivalent to Fig.~\ref{rombo3} (b), {\it i.e.} the same diagram as in Fig.~\ref{handle1}(b). 

The message is clear: there is {\it only one class of  diagrams that develop a tetraquark pole.}
It is  possible that  Figs.~\ref{handle1}(b) and \ref{rombo3}(b) admit tetraquark poles in $s$- and $t$- channels but not in their $u$-channel, corresponding to the cut in Fig.~\ref{rombo3}(a). The situation would be the same of meson-meson scattering in leading order, see~\cite{Witten:1979kh}.

\paragraph{{\bf One flavor and one antiflavor equal}.}
This case has been considered in~\citep{knecht} restricted to planar diagrams. Here we consider the realistic case where the tetraquark pole arises at the level of non-planar diagrams.~

The simplest case is given in  Fig.~\ref{manypoles}(b), corresponding to the tetraquark $b u\bar c\bar u$~\citep{Lucha:2017mof}. The coupling $g$ and the mixing parameter $f$ are found in the previous paragraph. From the figure, one derives
\be
f g \frac{1}{\sqrt{N}}=\frac{1}{N^4}, \nonumber
\ee
consistent with Eqs.~(\ref{couplingfin}) and~(\ref{mixing}).

\section{Zweig rule, tetraquarks and charmonium decays}  
Heavy quark pairs are not easily annihilated in hadronic transitions. This is the content of the Zweig rule, well obeyed from strange to beauty pairs: we may infer the presence of hidden charm or beauty from the presence of the heavy quarks in the final states, in open or hidden form.
If we neglect the diagrams with $c\bar c$ or $b\bar b$ annihilation, we may replace $b\bar c$ with  $c\bar c$ in the considerations of the previous paragraphs and apply the results to tetraquarks of composition $c u\bar c \bar q$, with $q=u,d$.

The two meson decay coupling is given by Eq.~(\ref{couplingfin}) and the mixing to $c\bar c$ charmonia by (\ref{mixing}). 

An interesting case is the decay $Y(4260)\to \mu^+ \mu^-$, which is implied by the direct production of $Y(4260)$ in  $e^+ e^-$ annihilation~\citep{Ablikim:2013dyn}. This decay cannot occur to lowest order in $\alpha$ via the irreducible diagrams since the e.m. current can annihilate only one quark-antiquark pair~\cite{Chen:2015dig}.  However, by (\ref{mixing}), $Y(4260)$ may decay via its mixing to the expected (but not yet identified) $L=2, S=J=1$ charmonium. This is a reason to prefer the new solution we have studied here, to Option 2 where tetraquark pole is in Fig.~\ref{handle1}(a) only. 

\begin{figure}
\begin{center}
   \includegraphics[width=0.90\columnwidth]{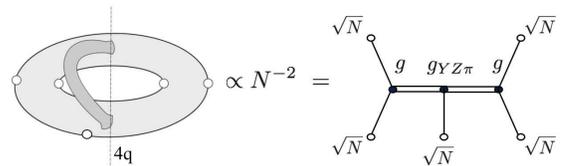}
 \end{center}
\caption{\footnotesize Diagram to describe the transition of a tetraquark into a lower tetraquark with pion emission, {\it e.g.} $Y\to Z^+ \pi^-$, see Eq.~(\ref{yzp}).}   
  
\label{deexc}
\end{figure}

One can consider the de-excitation of a neutral tetraquark, e.g. $Y\to Z^+ \pi^-$, by inserting one additional quark bilinear in one of the fermion loops in Fig.~\ref{luchamod}. Comparing with the expression in terms of tetraquark poles, Fig.~\ref{deexc}, one finds
\be 
 g_{_{YZ\pi}} \propto \frac{1}{\sqrt{N}}.
 \label{yzp}
 \ee

Replacing the meson insertion with the electromagnetic current, one obtains the amplitude for radiative decays, such as $Y(4260) \to X(3872) +\gamma$, reported in~\citep{Ablikim:2013dyn}. There is no $\sqrt{N}$ normalization factor for the current and the radiative decay rates are of order $\alpha N^0$.

\section{Summary and Conclusions}

We have addressed the problem of the lowest order in the $1/N$ expansion of QCD where a tetraquark pole may appear. In the more recent literature, the issue was considered in Refs.~\cite{Maiani:2016hxw} and~\cite{Lucha:2017mof}, see the Introduction of this paper for earlier references. 

We argue that the diagram of order $N^0$ proposed in Ref.~\cite{Lucha:2017mof},  Fig.~\ref{lucha1}(b) in the present paper, does not necessarily develop a tetraquark pole since the four-quark cuts present there may correspond to two non-interacting $q\bar q$ pairs. Introducing the lowest order interaction between the pairs, Figs.~\ref{handle1}(b) or \ref{luchamod}(c), brings the order of the diagram to $N^{-2}$. Saturation with tetraquarks of diagram~\ref{luchamod}(c)  raises a problem of compatibility with the similar saturation of the other diagram proposed in Reffs.~\cite{Maiani:2016hxw,Lucha:2017mof}, Figs.~\ref{handle1}(a) and \ref{lucha1}(a). The inconsistency cannot be solved by introducing two (or more) tetraquarks for a given flavor, as proposed in~\cite{Lucha:2017mof} for their correlation functions.  Rather, we conclude that only one of the two classes of diagrams will develop a tetraquark pole, either  Figs.~\ref{handle1}(b) and \ref{luchamod}(c), or Figs.~\ref{handle1}(a) and ~\ref{lucha1}(a).  

 {\it{\bf Option 1}} leads to the solution illustrated in this paper and, in particular, it allows for one tetraquark for given flavour distribution. {\it {\bf  Option 2}} gives back the solution found in Ref.~\cite{Maiani:2016hxw} for tetraquark-meson-meson coupling at the same time suppressing the mixing of tetraquarks with charmonia. We tend to disfavour this possibility, in view of the coupling of putative tetraquarks, the $Y$-states with $J^{PC}=1^{--}$, to the $e^+e^-$ channel.


Our analysis shows that nonplanar interactions are a {\it necessary condition} for the existence of tetraquarks in four-quark correlators. Assuming this to be the case, our solution is consistent with one tetraquark for given flavour and it includes tetraquark-charmonium mixing. 

Decay amplitudes of the states in meson-meson (but not the deexcitation of a higher tetraquark) will be decreased, in the $1/N$ expansion, with respect to previous, planar interaction, analyses, but it is not clear to us if this has real phenomenological implications. 

The existence of genuine, double heavy, tetraquarks studied in~\cite{Esposito:2013fma,Guerrieri:2014nxa} and in~\cite{Karliner:2017qjm,Eichten:2017ffp} is supported by independent theoretical approaches. If so, they must  appear in some order of the $1/N$ expansion in QCD: here we have discussed the necessary conditions for this to occur. 
\newline\newline
We acknowledge useful correspondence with 
R. Lebed.

\end{document}